\documentclass[prd,showpacs]{revtex4}
\usepackage[dvipdfm]{hyperref,graphicx}
\usepackage{bm}
\begin{document}
\title{Explanation to the Problem of the Naturalness measurements}

\author{Su Yan}
\email{yans@feynman.phys.northwestern.edu}
\affiliation{Department of Physics and Astronomy,
Northwestern University, Evanston, IL 60208, USA}
\begin{abstract}
The sensitivity parameter is widely used in measuring the severity of 
fine-tuning, while many examples show it doesn't work under certain circumstances. 
The validity of the sensitivity is in question. 
We argue that the dimensional effect is the reason why it fails in those scenarios.
To guarantee the sensitivity parameter correctly reflects the severity of fine-tuning,
it should be avoided to use under these specific circumstances.  
\end{abstract}

\maketitle

\section{Introduction}
The principle of naturalness introduced by Wilson and 't Hooft\cite{Wilson} requires 
the radiative corrections to a measurable parameter should not be much larger
than the measurable parameter itself. Therefore we don't need a magical fine-tune mechanism
that can precisely adjust the parameters of a theory.

The simplest example of a naturalness problem is the quadratically divergence of 
the fundamental scalar of \(\phi^4\) model:

\begin{equation}
\mathcal{L}=\frac{1}{2}[(\partial_{\mu}\phi)^2-m^2_0\phi^2]-\frac{g}{4!}\phi^4
\label{eqx1}
\end{equation}

At one-loop the renormalization of a scalar mass is of the form:

\begin{equation}
m^2=m^2_0-g^2\Lambda^2
\label{eqx2}
\end{equation}
where \(m_0\) is the bare mass, \(\Lambda\) is the cut-off energy scale. 
Because of the tremendous energy scale difference between the light scalar mass and the 
bare mass, 
a fine-tuning mechanism that can adjust \(m_0\) and \(\Lambda\) 
very precisely is required here. Otherwise any minute variation of \(m_0\) or \(\Lambda\) 
will completely change the value of the light scalar mass. 

In order to qualitatively describe the severity of fine-tuning, 
R. Barbirei and G.F.Giudice et al.\cite{BG} proposed a sensitivity parameter \(c\)
to  calculate the severity of fine-tuning.
If we have a fundamental Lagrangian parameter \(x\) and a measurable parameter \(y\), 
If we varies the Lagrangian parameter \(x\),  based on 
the corresponding variation of the observable parameter \(y\), 
the sensitivity parameter \(c\) is defined as:

\begin{equation}
c(x_0)=\bigg| \frac{x}{y}\frac{\partial y}{\partial x}\bigg|_{x=x_0}
\label{eqx4}
\end{equation} 

Here we need to emphasis that  the Lagrangian parameter \(x\) and the 
measurable parameter \(y\) may have different mass dimensions, while the sensitivity parameter \(c\)
is used to compare the fine-tuning properties of different models. 
We would like to see it is universal and model independent. A dimensionless 
sensitivity parameter is certainly the best choice. So in the definition of the sensitivity parameter 
\(c\), relative variations \(\delta y/y\) and \(\delta x/x\), rather than \(y\) and \(x\) 
are used as the basis of comparison. 
With this definition, larger sensitivity means higher severity of fine-tuning. 
In order to judge whether a parameter is fine-tuned or not, people usually
choose sensitivity \(c=10\) as the maximum allowed sensitivity for any
parameter to be categorized as ``natural''.  Any model or any parameter with 
\(c\gg 10\) is ``unnatural'' (or fine-tuned), and should be abandoned, although 
the definition of the cut-off sensitivity \(c=10\) is quite arbitrary.

The sensitivity criterion has been subsequently adopted by many researchers since then.  
It has been applied to many fields. Although in many examples the largeness of 
the sensitivity parameter is usually in good
correspondence to the fine-tuning, many researchers soon found it does not
accurately represent the severity of fine-tuning in certain scenarios. 
The most famous examples among them are examples given by G. Anderson et al\cite{GWA} and P. Ciafaloni et al\cite{CS}.

The example given by  G. Anderson et al\cite{GWA} is regarding the high sensitivity
of proton mass \(m_p\)  to the variation of the strong coupling constant \(g\).
Because the relation between the proton mass \(m_p\) and the strong coupling constant \(g\)
and the Planck scale \(M_P\) is:
\begin{equation}
m_p \approx M_P\exp{\Biglb[-\frac{(4\pi)^2}{bg^2(M_P)}\Biglb]}
\label{eqx5}
\end{equation}
which yields:
\begin{equation}
c(g)=\frac{4\pi}{b}\frac{1}{\alpha_s(M_P)}\gtrsim 100
\label{eqx6}
\end{equation}

According to the sensitivity criterion, the proton mass is definitely fine-tuned. But it is well known that
the lightness of the proton mass is the result of the gauge symmetry, not the result
of any fine-tuning mechanisms. Here the sensitivity parameter failed to reflect
the severity of fine-tuning correctly.

The example given by P. Ciafaloni et al\cite{CS} is about the high sensitivity of 
the Z-boson mass. When the Z-boson mass 
\(M_Z\) is dynamically determined through gaugino condensation in a ``hidden'' sector, 
the mass \(M_Z\) can be expressed as:
\begin{equation}
M_Z\approx M_P e^{-l/g^2_H}
\label{eqx7}
\end{equation}
where \(g_H\) is the hidden sector gauge coupling constant renormalized at \(M_p\),
and \(l\) is a constant.  Like the first example, the calculated sensitivity \(c\) for 
the second example is always much larger than the maximum allowed sensitivity.
If we follows the sensitivity criterion, the conclusion is
Z-boson mass \(M_Z\) is always fine-tuned. This is definitely not true. 

Actually, if we calculate the sensitivity \(c\) of any weak scale mass 
by varying a related coupling constant at the grand unification scale, the sensitivity
parameters are always very large no matter whether the weak scale mass is really
fine-tuned or not. All these examples show the sensitivity parameter is not reliable.

If we compare Eq.~(\ref{eqx5}) with Eq.~(\ref{eqx7}), it is easy to find both of 
them have a very similar mathematical formation. This gives us a hint that the sensitivity
parameter fails to reflects the severity of fine-tuning in these examples may be related
to this specific mathematical formation. The exponential function in Eq.~\ref{eqx5} and 
Eq.~\ref{eqx7} implies these problem may relate with the mass dimensions of these parameters.

To solve these problems, many authors attempted to explain this problem and proposed 
alternative prescriptions that supposed to be able to give correct results under these circumstances.
G. Anderson et al~\cite{GWA,AC2,AC3}. first introduced the idea of probability
distribution. They argued that, some physical parameters do have intrinsic large
sensitivity. We need to use \(\bar{c}\), the probability average of the 
sensitivity \(c\), to rescale the sensitivity parameter, the result will reflect the
fine-tuning level correctly:

\begin{equation}
\gamma = c/\bar{c}
\label{eqx8}
\end{equation}

Based on this criterion, only those with \(\gamma \gg 1\) can be considered as fine-tuned.
But they didn't explain why some parameters may have intrinsic large sensitivity and why
the rescaling can remove the problem. 

Other authors also proposed another modified version of the sensitivity parameter to solve this 
problem~\cite{CS,BR,BS,GRS,RS}:

\begin{equation}
c(x_0)=\bigg| \frac{\Delta x}{y}\frac{\partial y}{\partial x}\bigg|_{x=x_0}
\label{eqx9}
\end{equation} 
where \(\Delta y\) is the experimentally allowed range of the parameter \(x\). They argued
that correctly choose the experimentally allowed range would solve these problems. But as 
we know, the fine-tuning problem is an intrinsic property of a given model. It is nothing
to do with how we measure a physical quantity experimentally. 

Although many authors attempted to give a correct numerical description of the naturalness
level, yet none of them can claim quantitative rigor. No explanation has ever been proposed
to explain why sometimes we have such a large sensitivity for a not fine-tuned parameter.
Because it is still unclear how to quantitatively describe the fine-tuning problem correctly. 
The calculated fine-tuning level usually depends on what criterion we use, and how we use it.
Their judgments may reflect the naturalness properties correctly or incorrectly.
Because the sensitivity criterion plays such an important role, it is worth to investigate 
the relationship between the naturalness and the sensitivity, find the reason why
the sensitivity is so large for the examples we just discussed.

\section{The effect of different dimension}

It is meaningless to directly compare two physical quantities with completely different mass dimensions. 
They first need to be converted to a comparable format. 
Converting two different parameters \(x\) and \(y\) to the dimensionless formations 
\(\delta x/x\) and  \(\delta y/y\) is the method we
used in the definition of the sensitivity parameter. 
It looks the problem of comparing parameters with different mass dimensions has been solved,
but later we will find some side effects introduced by different mass dimension are still there.

To study these possible side effects, suppose we have a model with a dimensionless parameter \(x\)
(for example, a gauge coupling constant) and two dimensional parameters \(y\) and \(z\) 
(for example, two masses). We choose this specific model is because usually when we calculate 
the sensitivity parameter of a model, we either compare a mass with a coupling constant(for example,
Eq.~\ref{eqx6}), or compare 
a mass with another mass(for example, Eq.~\ref{eqx2}). The problems of the sensitivity parameter discussed 
in the first section
implied that it could be originated from comparing a dimensional parameter 
with a dimensionless parameter. To investigate all possibilities, 
we will discuss this issue into two parts: Comparing
a dimensional parameter with a dimensionless parameter, and comparing a dimensional 
parameter with another dimensional parameter.

\subsection{Comparing a dimensional parameter with a dimensionless parameter}

Generally, for a dimensional parameter \(y\) and a dimensionless parameter \(x\)
of a general gauge theory model, their corresponding lowest order renormalization group equations are\cite{Luo}:

\begin{equation}
\frac{dy}{dt}=\gamma(x)y+\cdots
\label{rge-a1}
\end{equation}

\begin{equation}
\frac{dx}{dt}=\beta x^3+\cdots
\label{rge-a2}
\end{equation}
where \(t=\ln Q/Q_0\), \(Q\) is the renormalization energy scale, \(Q_0\) is the grand unification scale. 
\(\beta\) and \(\gamma\) are dimensionless coefficients. Although for general gauge theories, the lowest
order term of the renormalization group equation of the scalar quartic coupling 
could be the first order term\cite{Luo}, 
generally, for a four dimensional problem, the lowest order term for other dimensionless couplings(for example, gauge couplings 
and Yukawa couplings) are the third order term\cite{Luo}. Also because the difference of the exponent
actually won't change the final result, so without loss of generality, to simplify the calculation here 
we assume the lowest order term of Eq.~\ref{rge-a2} is the third order term.

Because here \(y\) is a dimensional parameter, while \(x\) is a dimensionless parameter, 
thus for the right hand side of Eq.~\ref{rge-a1},  \(\gamma(x)y+...\) is the only possible
formation of the lowest order term. Otherwise each term will not have a consistent mass
dimension. Here  \(\beta\) is a constant decided by the specific unification model
while \(\gamma\) is a function of dimensionless coupling constants\cite{Luo,MV1,MV2} . 

At the grand unification scale \(t_0\), if given the initial conditions 
\(x=x_0\) and \(y=y_0\),  the solutions of Eq.~\ref{rge-a1} and Eq.~\ref{rge-a2}
are:

\begin{equation}
y=y_0 e^{\int_{t_0}^{t} \gamma(x) dt}
\label{equ-y}
\end{equation}

\begin{equation}
x^2=\frac{x_0^2}{1-2x_0^2\int_{t_0}^{t}\beta dt}
\end{equation}

According to Eq.~\ref{eqx4}, the sensitivity parameter of the dimensional parameter \(y\) to the variation of the 
dimensionless parameter \(x\) is:

\begin{equation}
c=\bigg|\frac{\partial \ln y}{\partial \ln x}\bigg|_{x=x_0}=
\bigg|\frac{\partial}{\partial \ln x} \int_{t_0}^{t}\gamma(x) dt\bigg|_{x=x_0}
\label{equ-c1}
\end{equation}

For general gauge theories, 
\(\gamma\) function is derived from the radiative correction to the 
propagator. Based on possible interactions, the coefficient \(\gamma(x)\) in Eq.~\ref{rge-a1}
generally takes the form of \(\gamma(x)=bx^2\) (\(b\) is a constant decided by the gauge group and
the interaction)\cite{Luo,MV1,MV2}. 
At the weak scale, most coupling constants are around the order of 
magnitude of 1\cite{pdg}, while corresponding to the Planck scale \(M_p=2.4\times 10^{18}\) GeV, 
the energy scale parameter is \(t_0\approx 36\). So roughly estimate the sensitivity 
parameter \(c\) (Eq.~\ref{equ-c1}) will be around 40.  According to the sensitivity criterion, the 
dimensional parameter \(y\) will be categorized as fine-tuned, even though it may not be. 
The sensitivity of Eq.~\ref{equ-c1} is small only when \(\gamma\Delta t \ll 1\). 
Consider the fact that \(\Delta t\) is around 40, it is quite unusual to have a small sensitivity.

From the dimensional analysis point of view, this is because these two parameters
have different mass dimensions. One is a 
dimensional while the other is dimensionless.  The
consistency of the mass dimension requires that these two parameters can not 
be coupled arbitrarily in the right hand side of the corresponding  renormalization group equations. 
If we integrate these two
renormalization group equations, and express the dimensional parameter as the 
function of the dimensionless parameter, the result
will be an exponential function. This result is consistent with Eq.~\ref{eqx5} and Eq.~\ref{eqx7}.
The consequence of this is these two parameters will have native 
order difference.  Small variation
of a dimensionless parameter always results in a big variation of the coupled 
dimensional parameter. Therefore
the calculated sensitivity usually will always be extremely large no matter whether it is really fine-tuned or not. 
Just like many authors have already pointed out, using the sensitivity parameter here 
will overestimate the severity of fine-tuning.

\subsection{Comparing a dimensional parameter with another dimensional parameter}

Calculate the sensitivity by comparing a dimensional parameter with another dimensional
parameter is another way used to 
estimate the severity of fine-tuning. Now suppose we have a dimensional parameter \(y\) 
and another dimensional parameter \(z\). Because of the mass dimension, their
lowest order renormalization group equations should be in the form of:

\begin{equation}
\frac{dy}{dt}=b_{11}y + b_{12}z+\cdots
\label{rge-b1}
\end{equation}

\begin{equation}
\frac{dz}{dt}= b_{21}z + b_{22}y+\cdots
\label{rge-b2}
\end{equation}
where coefficients \(b_{ij}\) are dimensionless and are functions 
of dimensionless coupling constants.

If the initial conditions at a specific energy scale \(t=t_0\) are 
 \(y=y_0\) and \(z=z_0\). For convenience, define the coefficient matrix \(B\) as:

\begin{equation}
B =\left( \begin{array}{cc}
b_{11}, b_{12} \\
b_{21},b_{22} 
\end{array}\right)
\end{equation}

The original renormalization group equations now can be rewritten as:
\begin{equation}
\frac{d}{dt}\left( \begin{array}{c}
y \\
z
\end{array} \right)
=B \left( \begin{array}{c}
y \\
z
\end{array} \right)
\label{matrix}
\end{equation}

Because the couplings are functions of \(t\), so the matrix \(B\) is also a
function of \(t\). This means Eq.~\ref{matrix} won't have
a closed-form solution. Its solution usually
can be written as a Magnus series\cite{magnus}:
\begin{equation}
\left( \begin{array}{c}
y \\
z
\end{array} \right)
=\left( \begin{array}{c}
y_0 \\
z_0
\end{array} \right)e^{\tilde{B}t}
\label{matrix2}
\end{equation}
where \(\tilde{B}\) can be expressed as a Magnus series:
\begin{equation}
\tilde{B}=\frac{1}{t}
\Biglb(\int_{t_0}^{t}B(t)dt-\frac{1}{2}\int_{t_0}^{t}\int_{t_0}^{\tau_1}[B(\tau_2),B(\tau_1)]d\tau_2d\tau_1+\cdots\Biglb)
\end{equation}

Suppose the eigenvalues of the matrix \(\tilde{B}\) are \(\lambda_1\) and \(\lambda_2\), their corresponding eigenvectors are
\(\bm{\eta_1}\) and \(\bm{\eta_2}\),  rewrite Eq.~\ref{matrix2}: 

\begin{equation}
\left( \begin{array}{c}
y \\
z
\end{array} \right)
=c_1 e^{\lambda_1t}\bm{\eta_1}+c_2 e^{\lambda_2t}\bm{\eta_2}
\label{eqx20}
\end{equation}
where \(c_1\) and \(c_2\) are constants which will be decided by the initial conditions.

Because the coefficients \(b_{ij}\) \
are functions of the coupling constants, so the matrix \(\tilde{B}\) is the function of
the couplings and the energy scale \(t\). Because
\(\lambda_i\) and \(\bm{\eta}_i\) are eigenvalues and eigenvectors of the matrix \(\tilde{B}\),
therefore  \(\lambda_1t\) and \(\lambda_2t\),  
\(\bm{\eta_1}\) and \(\bm{\eta_2}\) are all functions of the coupling
constants and the energy scale \(t\). 
 
Take the initial conditions \(y=y_0\) and \(x=x_0\) into account, 
Eq.~\ref{eqx20} can be rewritten in the following explicit form:
\begin{equation}
y=b_1(y_0,z_0)e^{\lambda_1t}+b_2(y_0,z_0)e^{\lambda_2t}
\label{eqe-y}
\end{equation}
\begin{equation}
z=b_3(y_0,z_0)e^{\lambda_1t}+b_4(y_0,z_0)e^{\lambda_2t}
\label{rqe-z}
\end{equation}

Where the coefficients \(b_1,b_2,b_3,b_4\) have the same mass dimension as \(y\) and \(z\).
They are functions of the initial values \(y_0\) and 
\(z_0\), and the exponents \(\lambda_1\) and \(\lambda_2\) are explicit functions of the coupling 
constants.

According to Eq.~\ref{eqx9}, the corresponding sensitivity parameter is:

\begin{equation}
c=\bigg|\frac{\partial \ln y}{\partial \ln z}\bigg|_{z=z_0}=\bigg|\frac{\frac{\partial}{\partial ln z_0}(b_1
e^{\lambda_1t})+\frac{\partial}{\partial ln z_0}(b_2 e^{\lambda_2t})}
       {b_1e^{\lambda_1t}+b_2e^{\lambda_2t}}\label{equ-c2}\bigg|
\label{eqx23}
\end{equation}

In Eq.~\ref{eqx23}, if either the numerator or the denominator dominants,
the calculated sensitivity will be either very large or very small intrinsically,
the sensitivity criterion may not give a correct result. To rule out this possibility,
We need to further estimate Eq.~\ref{eqx23} and need to know the value of 
\(\partial (b_i\exp(\lambda_i t))/\partial \ln z_0\), which means we 
need to know how the coupling constants respond to the variation
of a mass parameter.  

As we know, only very limited numbers of parameters are considered
as independent and fundamental Lagrangian parameters. The remaining parameters 
need to be calculated. Certainly the gauge coupling constants are 
independent of the variation of any mass parameters, but the Yukawa couplings are different.
They need to be calculated from the 
fermion masses and Vevs. Depends on which mass parameter we choose to measure the
sensitivity, sometimes they are not independent. This means
the exponents \(\lambda_1\) and \(\lambda_2\) could be 
implicit functions of the mass parameter we vary.   
 
Because Yukawa couplings are dimensionless parameters, the lowest order terms of
their renormalization group equations will not contain any mass parameter. So a Yukawa
coupling constant won't be very sensitive to
the variation of a fundamental mass parameter.

Thus  the derivative:

\begin{equation}
\frac{\partial}{\partial\ln z_0}(b_i e^{\lambda_i t})=
e^{\lambda_it}(z_0\frac{\partial b_i}{\partial z_0})+b_i e^{\lambda_i t}t\frac{\partial \lambda_i}{\partial \ln z_0}
\end{equation}

The second term usually is small; while for the first term, 
because \(b_i\) has  the same mass dimension as \(z_0\),
depends on which mass parameters we are comparing, and 
depends on the specific Lagrangian and specific values of these
parameters, the calculated  sensitivity could be either small or large.
It won't have the same problem discussed in part A. Therefore if two parameters have the 
identical mass dimension, the sensitivity parameter
can reflect the severity of fine-tuning correctly.

\section{Effects on the validity of the sensitivity Criterion}

The discussion in section II shows that, in a general gauge
theory, due to the requirement of 
the mass dimension consistency in the renormalization group equations, usually the
mathematical relation between
a dimensionless parameter at one energy scale and a dimensional parameter at
another energy scale is an exponential
function. A dimensional parameter is always  sensitivity to the 
variation of a dimensionless parameter, even though they are not fine-tuned. 
As a result, the severity of fine-tuning will be overestimated if using the sensitivity
parameter under this specific condition.
This is the reason why the sensitivity criterion failed in the examples discussed
in section I. The discussion in section II also shows when comparing parameters 
with identical mass dimensions, we won't have the same problem. 

On the other hand, although renormalization is important for the
fine-tuning, the naturalness problems also exist in problems 
not related to the renormalization\cite{AD}. 
While the effects introduced by different mass dimensions only exists when 
two parameters compared are in different energy scales, which means the sensitivity
parameter won't have this problem in the naturalness problems not related to the renormalization.
For instance, one is a weak scale observable
parameter while the other is a Lagrangian parameter at the grand unification scale. 
These two parameters need to be linked by renormalization, 
therefore the dimensional effect plays an important role. 
If all parameters involved in a fine-tuning problem are at the same energy scale,
for example, the problems related to the mass mixing mechanisms\cite{Casas:2004gh,AD}, 
a renormalization mechanism is not required to link these two parameters. Thus
the dimensional effect has no place to play a role under such 
conditions. The severity
of fine-tuning won't be overestimated if using the sensitivity parameter.

Therefore, the accuracy of the sensitivity criterion can not
be guaranteed only when  two parameters with different 
mass dimensions at different energy scales are compared. Because 
two parameters at different energy scales need to be connected by the renormalization relations. 
The effect introduced by the different mass dimension thus can play an important role
in the measuring of the fine-tuning level. 

Many alternative methods have been proposed since the problem of the sensitivity
parameter was uncovered\cite{GWA,AC2,AC3,CS,BR,BS,GRS,RS}. These methods 
restrict the sensitivity parameter either by
presetting an upper limit to the value of one language parameter or 
rescaling the sensitivity calculated by the background sensitivity. 
These alternative prescriptions 
can relieve the problem, but technically, in order to solve the
problems introduced by different mass dimensions,  we need to know how the dimensional 
effect contributes to the sensitivity. In other words, how to properly calculate the native 
background sensitivity used to remove the contributions other than the fine-tuning. 
 
We know unlike a free field,  for a real model, many mechanisms contribute to 
the dimension of a parameter. The engineering dimension usually is not equal to the canonical
dimension. Besides, there are quantum corrections. How much each mechanism contributes
depends on the specific model and specific values of these parameters. 
It is difficult and not efficient to invent a simple and universal prescription 
that can be easily applied to any models for any situations.

As a solution, it is better to avoid using the sensitivity to compare parameters with different mass dimensions
if these parameters are at the different energy scales and need to be related by
renormalization.
Because this is the most simple and most efficient way to measure the naturalness correctly, 
and in the same time, avoid the problems introduced by the difference mass dimensions.
Besides, avoid comparing parameters with different mass dimension won't restrict
our ability to describe the fine-tuning properties, because we can always measure the 
severity of fine-tuning by comparing parameters with same mass dimension.

\section{Conclusion}
The problem of the sensitivity parameter has been discovered more than ten years.
Many analyses, explanations and alternative prescriptions have been proposed to 
solve this problem. But none of them have been widely accepted. the reason why the
sensitivity parameter fails under certain circumstances so far still remains unclear.
In this paper we have investigated the problems existed in the 
sensitivity parameter, found the reason why the sensitivity parameter failed to 
represent the true level of fine-tuning is because of comparing parameters with 
different mass dimension at different energy scale, which introduced the dimensional effect.
As a consequence, the sensitivity parameter will overestimate the severity of 
fine-tuning, can not reflect the fine-tuning properties correctly. Certainly,
this effect only exists when parameters with different 
mass dimensions are compared  at different energy scales.
The best way to avoid this problem is always compare parameters with same mass dimensions
if they are at different energy scales.

\end{document}